\documentclass[prl,twocolumn,showpacs]{revtex4}

\begin{document}

\date{\today} 

\newcommand{\beq}{\begin{equation}}
\newcommand{\eeq}{\end{equation}}
\def\d{{\delta}}
\def\ba{\begin{array}}
\def\ea{\end{array}}
\def\be{\begin{equation}\begin{array}{l}}
\def\ee{\end{array}\end{equation}}
\def\bea{\begin{equation}\begin{array}{l}}
\def\eea{\end{array}\end{equation}}
\def\f#1#2{\frac{\displaystyle #1}{\displaystyle #2}}
\def\om{\omega}
\def\omm{\omega^a_b}
\def\we{\wedge}
\def\de{\delta}
\def\De{\Delta}
\def\va{\varepsilon}
\def\omb{\bar{\omega}}
\def\la{\lambda}
\def\vv{\f{V}{\la^d}}
\def\si{\sigma}
\def\t{T_+}
\def\v{v_{cl}}
\def\m{m_{cl}}
\def\n{N_{cl}}
\def\bi{\bibitem}
\def\c{\cite}
\def\sa{\sigma_{\alpha}}
\def\ua{\uparrow}
\def\da{\downarrow}
\def\mua{\mu_{\alpha}}
\def\ga{\gamma_{\alpha}}
\def\g{\gamma}
\def\G{\Gamma}
\def\ora{\overrightarrow}
\def\pa{\partial}
\def\ov{\ora{v}}
\def\al{\alpha}
\def\bt{\beta}
\def\R{R_{eff}}
\def\th{\theta}
\def\na{\nabla}

\def\muu{\f{\mu}{ed}}
\def\E{\f{edE(\tau)}{\om}}
\def\t{\tau}
\def\lam{\lambda}
\def\dd{d^{\dagger}}
\def\sb{{\bar{s}}}
\def\half{{1\over2}}
\def\third{{1\over3}}
\def\twof{{2\over5}}
\def\threes{{3\over7}}
\def\rhob{{\bar \rho}}
\def\ua{\uparrow}
\def\da{\downarrow}
\def\eqa{\begin{eqnarray}}
\def\eea{\end{eqnarray}}
\def\beqr{\begin{eqnarray}}
\def\eeqr{\end{eqnarray}}

\title{Coherence Network in the Quantum Hall Bilayer}

\author{H. A. Fertig$^1$ and Ganpathy Murthy$^2$}

\affiliation{1. Department of Physics, Indiana University, Bloomington, IN 47405\\
2. Department of Physics and Astronomy, University of Kentucky, Lexington, KY 40506-0055}

\begin{abstract}
Recent experiments on quantum Hall bilayers near total filling factor 1
have demonstrated that they support an ``imperfect'' two-dimensional 
superfluidity, in which there is nearly dissipationless transport
at non-vanishing temperature observed both in counterflow resistance
and interlayer tunneling.  We argue that this behavior may be understood
in terms of a {\it coherence network} induced in the bilayer by
disorder, in which an incompressible, coherent state exists in narrow
regions separating puddles of dense vortex-antivortex pairs.
A renormalization group analysis shows that it is appropriate to describe
the system as a vortex liquid.  We demonstrate that the dynamics of
the nodes of the network leads to a power law temperature dependence
of the tunneling resistance,
whereas thermally activated hops of vortices across the links
control the counterflow resistance.  
\end{abstract}

\maketitle

{\it Introduction}-- Over the past decade, there has been
accumulating evidence that excitonic superfluid-like \cite{keldysh} 
properties may be present in quantum Hall
bilayer systems \cite{bilayers} when the filling factor
of each layer is near $\nu=1/2$ and the layers are sufficiently close.
The relevance of excitons to this system may be understood by
using a filled Landau level in a single layer as
a starting reference state; equal densities in the layers may then 
be reached by creating a particle-hole condensate
\cite{fertig89,eisenstein}.  It has long been suspected that such a
state might have superfluid properties \cite{wen}.

The most dramatic superfluid-like properties of this system
are observed 
when the layers are separately contacted and a current
is passed from one layer to the other.
For contacts at opposite ends of the sample,
the resulting tunneling
conductance $\sigma_T$ has a sharp resonance precisely 
at zero voltage \cite{spielman1}, which is
reminiscent of the DC Josephson effect.
When the contacts are on the same side of the sample
and the layers on the opposite side are short-circuited,
so that currents in the two layers flow in opposite directions
[``counterflow'' (CF)],
the voltage drop in either layer along the direction of
flow tends to zero in the low temperature 
limit \cite{kellogg,tutuc}.  This may be
understood as nearly dissipationless
flow of electron-hole pairs.  Interestingly, the
temperature dependences of these two behaviors are very
different: the tunneling resistance $1/\sigma_T$ falls
much more slowly than the dissipative in-plane resistance. 

The ``imperfect'' superfluidity apparent in these experiments suggests
that a good starting point for understanding this system is in terms
of a condensed state with interlayer coherence.  However, a true
two-dimensional superfluid should have a power law $I-V$ in CF below
some critical Kosterlitz-Thouless temperature $T_{KT}$, whereas an
Ohmic response is observed at all available temperatures.  Moreover, a
true DC Josephson effect would yield an {\it infinite} tunneling
conductance below $T_{KT}$ rather than just a sharp resonance.  
It is also remarkable that in both the tunneling and CF geometries,
current appears to flow over the entire length of the system,
whereas for a true superfluid the current would decay within a
Josephson length of the edge \cite{prange}, which is far smaller.  The
linear response of the quantum Hall bilayer, exhibiting
dissipationless transport, if at all, only at zero temperature, is a
new transport regime for condensed matter systems.  Most theoretical
studies of this system suggest that disorder must be crucially
involved in the mechanism(s) that lead to this behavior
\cite{balents,hafjp2,wang,sheng}.

In this work, we describe a state where the experimentally observed
behavior emerges naturally, and which is indeed induced by the
disorder environment of the quantum Hall bilayer.  We will argue below
that the system breaks up into large regions in which the interlayer
coherence is small or absent, separated by narrow regions in which the
coherence is relatively strong.  The resulting structure is a {\it
coherence network}, with long, quasi-one-dimensional links connected
at nodes.  We will demonstrate that dissipation in tunneling may be
understood in terms of the phase dynamics of the nodes, whereas
dissipation in CF occurs due to thermally activated hops of vortices
across the links.  Further experimental consequences of our model are
discussed below.

{\it Nonlinear Screening and the Coherence Network} --
We start with the Efros model for nonlinear
screening of potential fluctuations in a quantum Hall system \cite{efros}.
This screening generates puddles of quasiholes or quasielectrons,
which are locally dense and mobile enough to destroy the incompressibility
of the quantum Hall fluid.
The puddles are separated by narrow ($\sim \ell_0$, with $\ell_0=\sqrt{\hbar c/eB}$
the magnetic length) incompressible
strips which follow the equipotentials of the total effective
potential, and within which there is a charge gap. The
strips form links which join together at nodes -- associated
with saddle points of the disorder potential -- to form a random network.

For the bilayer system, the incompressible state at total filling factor $\nu=1$
has a {\it phase} degree of freedom $\theta$ that may be understood
as the wavefunction phase for excitons in the incompressible
regions of the system.  The conjugate variable to this phase 
angle is the interlayer charge imbalance, which may be denoted as $S_z$.
For equal layer populations, on average $<S_z>=0$,
and the system becomes analogous to an easy-plane Heisenberg ferromagnet
\cite{moon}.  Vortices of the phase field (``merons'')
carry both an electric charge ($\pm e/2$ for either vorticity) and
an interlayer electric dipole moment \cite{moon,lee,huse}.
Thus, in this context the links and nodes of the network carry
a phase degree of freedom, while
the puddles are flooded with vortex-antivortex
pairs \cite{puddles}.  

The presence of meron-antimeron pairs at link edges
implies that
the phase angle 
$\theta({\bf r})$ turns over many times along a given link.
When tunneling is included, these  overturns
bear a close resemblance to kink solitons 
of the sine-Gordon model \cite{moon}.
As discussed below, the phase dynamics of the links is 
conveniently described in terms of these solitons.
Our picture of this coherence network is illustrated in Fig. 1.

{\it Renormalization Group Analysis} -- For counterflow experiments,
dissipation can only occur when merons -- i.e., vortices -- are
mobile and unbound \cite{huse}.  For this to occur, the vortices
in the coherence network must be in a {\it liquid} phase; i.e., their
discreteness should not inhibit their ability to screen 
``charge'' (i.e., enforced vorticity) at long distances.
We begin with a classical model on a square lattice network,
\begin{eqnarray}
H_{CN} = \sum_{\bf r} \Bigl\lbrace
{1 \over 2} K \sum_{\mu=x,y} 
\biggl[2 \pi m_{\mu}({\bf r}) - \partial_{\mu} \theta({\bf r})
+A_{\mu}({\bf r}) \biggr]^2 
- h \cos \theta({\bf r})
\Bigr\rbrace,
\nonumber 
\end{eqnarray}
where $m_{x,y}({\bf r})$ are integer
degrees of freedom representing the number of solitons on the
link extending  from the node at ${\bf r}$ in the $x$ or $y$ direction, and
$A_{\mu}({\bf r})$ is a Gaussian random variable obeying
$<A_{\mu}({\bf r}) A_{\nu}({\bf r^{\prime}})> = \Delta \delta_{\mu,\nu}
\delta_{{\bf r},{\bf r^{\prime}}}$
which models the effect of
disorder.  
The parameter $h$ is
proportional to the interlayer tunneling matrix element, which explicitly
breaks the $U(1)$ symmetry
in energetically favoring $\theta=0$.
 
It is convenient to recast the problem explicitly in terms of
vortices, which can be done by standard techniques \cite{jose},
leading to $Z=\int{\cal D}\phi \sum_{\lbrace B \rbrace} e^{-H_V}$,
with
\begin{eqnarray}
H_V=2\pi^2K \sum_{{\bf r}}
\Bigl\lbrace
|{\bf \nabla} \phi({\bf r}) + \hat{x} B({\bf r})
+ {\bf A}({\bf r}) |^2 \nonumber \\
-y_h \cos 2\pi \phi({\bf r}) 
+E_c \Bigl({{\partial B({\bf r})} \over {\partial y}} \Bigr)^2
\Bigr\rbrace.
\label{hv}
\end{eqnarray}
In this representation $B$ is an integer field that resides on
vertical bonds, and the vorticity is $m({\bf r})=\partial
B/\partial y$. Finally, $E_c$ represents a core energy for these
vortices, and $y_h=e^{-k_BT/h}$. This form of $H_V$ is appropriate
when $y_h <<1$, which is easily satisfied under experimental
conditions.  For $y_h=0$ and ${\bf A}=0$, $\phi$ may be integrated out
to recover the usual Coulomb gas.

To facilitate the renormalization group (RG) analysis, we replace the
integer field $B({\bf r})$ with a continuous field $b({\bf r})$ and
add a term $-y_b \sum_{\bf r} \cos[2\pi b({\bf r})]$ that tends to
keep $b$ at its intended integer values.  By matching low energy
configurations for the integer and continuous field versions of the
theory, one may show $y_b \sim E_c$ for small $E_c$ and $y_h$. The
crucial point is that {\it the irrelevance of $y_b \sim E_c$ will
indicate a liquid state of vortices}.

In performing the RG, $b$ is rescaled according to $b^{\prime}({\bf r}^{\prime})e^{-\ell}=
b({\bf r})$, where $e^{-\ell}$ is the rescaling factor,
in order to keep the Gaussian part of the Hamiltonian of fixed
form. As $b$ shrinks with rescaling, it is natural
to expand $\cos[2\pi b({\bf r})]$ in a power series in $b$,
and treat the fourth and higher order terms perturbatively.  
The quadratic term, however, introduces a contribution to the fixed point
of the
form ${1 \over 2} r \sum _{\bf r} b({\bf r})^2$. A key observation
is that this term 
limits large separations for vortices, so that if $r \ne0$ at
the fixed point, the system is in a bound vortex 
state \cite{deconfine1,deconfine2}.  In the clean limit (${\bf A}=0$),
it has been demonstrated that this happens
if the temperature is too
low, $E_c$ is too large, or if $h$ is too large.
By contrast, for $r=0$, states of arbitrarily large ``vortex dipole moment''
may be found in the thermodynamic limit, so that the vortices are
effectively unbound \cite{deconfine2}. 

To generalize this to the disordered case,
we use the replica trick before proceeding with the RG analysis.
In the $n \rightarrow 0$ limit, the flow equations become,
to lowest order in $E_c$ and $y_h$,

\begin{eqnarray}
{{d} \over {d\ell}}\Bigl( {{r} \over K} \Bigr) &=& 
- e^{-2\ell} \sqrt{{{r/K} \over {1+\xi^2\Lambda^2}}}\Lambda^2
\bigl[{1 \over K} +2\pi^2 \Delta \bigr] \nonumber\\
{{d \xi^2} \over {d\ell}} =  &-2\xi^2& \nonumber \quad
{{d y_h} \over {d\ell}} = (2-\eta/\sqrt{Kr})y_h \nonumber
\end{eqnarray}
where $\Delta$ is the disorder strength, $\Lambda=\pi/a_0$ is a
wavevector cutoff, $a_0$ is the distance between nodes, $\xi$ is a
length scale with initial value $\xi(\ell=0)=\sqrt{E_c/4\pi^2K}a_0$,
and the initial value of $r(\ell=0) \sim E_c$.  The parameter $\eta$
is a number of order unity.  To this order in perturbation theory, the
parameter $K$ remains constant.  For large disorder $\Delta$ and/or
small $K$ \cite{sheng,stiffness}, it is
apparent that $r , y_h \rightarrow 0$.

Since all signs of the discreteness of the underlying the vortex
density has scaled away at this fixed point, it is a liquid.  In this
phase neither the stiffness $K$ nor the symmetry-breaking field $h$
are effective at preventing phase overturns along a column of nodes
(say, at the edge of the system) if their phase angles are driven by
an arbitrarily small torque \cite{wei}.  Moreover, the irrelevance of
$y_h$ means the tunneling term $h \cos \theta$ may be treated
perturbatively in the vortex liquid phase.  We now exploit this
observation to understand the transport properties of the coherence
network.

{\it Dissipation in Transport: Tunneling and Counterflow} -- We can
understand transport in the coherence network based on the following
considerations. (i) The total current in the sample is the sum of a
CF current and the quantum Hall edge current. Dissipation
in the latter is negligible in both the tunneling and 
CF geometries we consider here, and so is ignored in what follows.
(ii) We assume the solitons are unpinned in the links, as is appropriated
for slowly varying disorder, and treat
them as an elastic system. The displacement field
$u(x)$ is related to the counterflow current via
$I_{CF}=\rho_s\partial_x\theta \approx {\tilde K}\partial_x u$ (with ${\tilde
K}\propto \rho_s$). From the Josephson relation, it follows that the
interlayer electric potential is $V_{int}={\hbar\over e} \partial_t\theta={h\over
e} \partial_{t} u/b_0$, where $b_0$ is the average 
spacing of the solitons. 
A simple effective Gaussian theory may be written for the
solitons on the links \cite{unpub}, whose form would control
dissipation at sufficiently high frequencies.  Such a purely
quadratic theory however cannot dissipate energy at zero frequency \cite{maslov-stone}. 
%
%
(iii) 
CF current can be
degraded by hopping of merons across the link. This is a thermally
activated process with energy $\Delta_{link}$, because
the merons must cross the barrier of the incompressible link. (iv) The
nodes  are
relatively meron-free regions with a high stiffness, and can be
treated as rigid rotors which are subject to torques $F_{link}\propto
I_{CF}^{link}$ from each attached link, and a force $-h\sin\theta$ due
to the tunneling.

Consider first the tunneling geometry. Since the current flows into
(say) the top layer on the left and leaves via the bottom layer on the
right, the CF current points in opposite directions at the two ends of
the sample. From (ii) above this implies the currents cause 
the underlying phase angles
in the links and nodes to rotate in the {\it same} direction,
and may be modeled by forcing in solitons at one sample end and removing
them at the other,
as illustrated in Fig. 2.  The 
dynamics of a typical node may be described
by a Langevin equation
\beq
\Gamma {d^2\theta\over dt^2}=\sum\limits_{links}F_{link}-\gamma_0 {d\theta\over dt} -h\sin{\theta}+\xi(t).
\eeq
Here $\Gamma$ is the effective moment of inertia of a rotor, proportional
to the capacitance of the node,
$\xi$ is a random (thermal) force, and $\gamma_0$ is the viscosity
due to dissipation from the other node rotors
in the system. For a small driving force, the node responds viscously,
and the resulting rotation rate 
determines the rate of flow of solitons via
$\gamma{\dot\theta}=\sum F_{link}$. From (ii) and
(iv) above one sees that the viscosity $\gamma$ is proportional to the tunneling {\it
conductance} $\sigma_T$ of the system. 
For $k_BT \gg h$ one may show the viscosity for an individual node to be \cite{dieterich}
\beq
\gamma=\gamma_0+\Delta\gamma=\gamma_0+\sqrt{{\pi\over{2\Gamma}}}{h^2\over (k_BT)^{3/2}}
\label{gammaequation}.\eeq
As each node contributes the same amount to the total viscosity,
the {\it total} response of the system to the injected CF current obeys
\beq
I_{CF}\propto N_{nodes} \Delta\gamma {e^2V_{int}\over\hbar}=\sigma_T V_{int}
\eeq
Note that because the nodes respond viscously, the tunneling conductance
is proportional to the area of the bilayer. This is so far
consistent with experiment but difficult to understand without the network
geometry posited here \cite{balents}.
The dependence of the tunneling conductance on temperature,
system area, and $h$ each constitute an experimentally
falsifiable prediction of the coherence network model of the
quantum Hall bilayer \cite{com1,com2}.

Now consider the CF geometry. This is 
created by short-circuiting the layers at one end of the
sample, which we will take to be on the right.
There the rotor angles should be
subject to free boundary conditions, leading to $F\propto{\dot
\theta}\propto{\dot u}=0$. The injected CF current on the left leads to a
compression of the solitons (or a rarefaction of antisolitons). 
In a uniform superfluid, such a force would be
fully balanced by the effective force due to tunneling, and the
CF current would flow without dissipation. 
In the coherence network, however, at any $T\ne0$, 
merons are able to thermally hop across the links to create or destroy
solitons, and the other (unpinned) solitons
move in response, leading to dissipation.

In equilibrium, the rates of creation and annihilation of
solitons in a link obey detailed balance.
However,
the presence of the ``force'' $F\propto I_{CF}$
destroys this balance, and leads to a net meron
current
\beq I_{meron}={\zeta\over e} I_{CF} e^{-\Delta_{link}/k_B T},
\label{cf-meron-relation}\eeq
where $\zeta$ is a dimensionless number depending on the details of
the link disorder.  If, for example, there is a net annihilation of
solitons with time on a given link, solitons will move in from the
left to take their place.
The rate of
change of $\theta$ due to such processes at a point inside the sample a distance $x$ from
the left end is 
\beq {d\theta(x)\over {d{\tau}}}=2 \pi
\int\limits_{x}^{L} I_{meron}(x') dx'
\label{thetadot-meron}.\eeq
For temperatures satisfying
$h \ll k_BT \ll \Delta_{link}$, the degrading of $I_{CF}$ due to
the nodes may be ignored.
In this case, Eqs. \ref{cf-meron-relation} and \ref{thetadot-meron} together
with the Josephson relation yield an activated longitudinal resistance measured 
between probes $\Delta x$ apart on a single layer
of the form
%
%
\beq
R_{xx}=  (\Delta x) \zeta {h\over e^2} e^{-\Delta_{link}/k_B T}
\eeq
which is consistent with experiment. 

In the CF geometry, the meron current $I_{meron}$ also creates a Hall voltage
within individual layers
because merons carry an interlayer dipole moment,  
leading to nonequilibrium charge imbalances at the upper and lower
edges of the sample of opposite sign.  
Assuming a $T$-independent resistive relaxation at the
edges (where the electrons are likely to be compressible),
this creates a Hall voltage of opposite sign on the
two layers which is activated, with the same activation energy as the
longitudinal resistance. This is consistent with measurements in
electron samples \cite{kellogg,holes}.

{\it Experimental Consequences} -- As shown in this work, the
imperfect superfluidity observed in quantum Hall bilayers may
be understood if we assume the electrons have organized into
a coherence network.
Several experimental tests of this hypothesis are possible.  
These include the temperature and size dependence
of the tunneling conductance discussed above. Another interesting
test would be measurement of the finite frequency tunneling
conductance, for which the response of the {\it links} should become
accessible.
An additional possibility is the introduction of an artificial
weak link across the direction of counterflow current.  This would
create a favored channel for soliton destruction/creation
that should lead to a
decrease in the activation energy for the CF resistance and an
enhanced Hall voltage at its endpoints.

The authors would like to thank S. Das Sarma, J.P. Eisenstein,
M. Shayegan, and E. Tutuc for useful discussions.
This work was supported by the NSF via Grant Nos.
DMR0454699 (HAF) and DMR0311761 (GM).



\begin{figure}
 \vbox to 5.0cm {\vss\hbox to 10cm
 {\hss\
   {\includegraphics{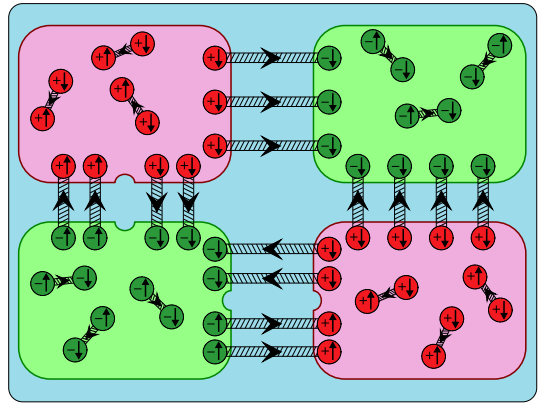}
   }
  \hss}
 }
\vspace{2mm}
\caption{Representation of coherence network.  Links and nodes
separate puddles of 
merons (circles).  Meron charge and
electric dipole moments indicated inside circles, as are
strings of overturned phase connecting meron-antimeron pairs.
}
\label{fig1}
\end{figure}

\begin{figure}
 \vbox to 4.0cm {\vss\hbox to 10cm
 {\hss\
   {\includegraphics{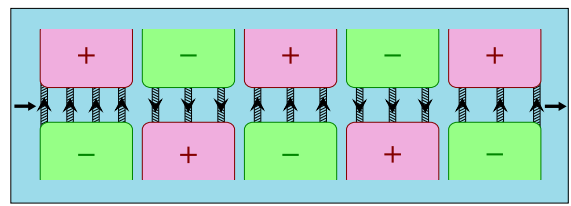}
   }
  \hss}
 }
\vspace{-14mm}
\caption{
Counterflow current in tunneling geometry, created
by adding solitons at one edge of sample and removing them at the other,
shown for a single set of links.
}
\label{fig2}
\end{figure}

\end{document}